\documentclass[aps,prl,reprint,superscriptaddress, longbibliography]{revtex4-2}

\usepackage []{inputenc}
\usepackage{siunitx}
\usepackage{amsmath,bm}
\usepackage{amsbsy}
\usepackage{amssymb}
\usepackage{mathptmx, textcomp}
\usepackage{xcolor}
\usepackage{braket}
\usepackage{graphicx}
\usepackage{textcomp}
\usepackage{notes2bib}
\usepackage{textcomp}
\usepackage{amsmath}
\definecolor{blouge}{rgb}{0.5, .1, .6}
\definecolor{bl}{rgb}{0, .1, .6}
\definecolor{turquoise}{rgb}{0.251, 0.878, 0.816}
\usepackage[colorlinks=true, citecolor = bl, linkcolor = bl, urlcolor=bl, pdfborder={0 0 0}]{hyperref}
\DeclareSIUnit\gauss{G}
\usepackage{placeins}
\let\Oldsection\section
\renewcommand{\section}{\FloatBarrier\Oldsection}

\usepackage[commentmarkup=uwave]{changes}

\newcommand{\bs}[1]{\boldsymbol{#1}}
\newcommand{\E}{\mathrm{e}}
\newcommand{\I}{\mathrm{i}}
\newcommand{\D}{\mathrm{d}}
\newcommand{\Szero}{{}^{1}\mathrm{S}_0}
\newcommand{\Pone}{{}^{3}\mathrm{P}_1}
\newcommand{\Done}{{}^{3}\mathrm{D}_1}
\newcommand{\ketbra}[2]{\ket{#1}\!\!\bra{#2}}

\begin{document}
\title{All-Optical Control of Birefringence in a Cold Atomic Ensemble}

\author{Apoorva Apoorva}
\affiliation{Université Côte d’Azur, CNRS, Institut de Physique de Nice, 06200 Nice, France}

\author{Naudson Lucas Lopes Matias}
\affiliation{Université Côte d’Azur, CNRS, Institut de Physique de Nice, 06200 Nice, France}

\author{Bérengère Pinoche}
\affiliation{Université Côte d’Azur, CNRS, Institut de Physique de Nice, 06200 Nice, France}

\author{Daniel Benedicto Orenes}
\affiliation{Université Côte d’Azur, CNRS, Institut de Physique de Nice, 06200 Nice, France}

\author{Robin Kaiser}
\affiliation{Université Côte d’Azur, CNRS, Institut de Physique de Nice, 06200 Nice, France}

\author{Raphaël Saint-Jalm}
\email[]{raphael.saint-jalm@univ-cotedazur.fr}
\affiliation{Université Côte d’Azur, CNRS, Institut de Physique de Nice, 06200 Nice, France}

\begin{abstract}
We demonstrate all-optical control of birefringence in a cold atomic cloud of ytterbium. By optically dressing the excited $\Pone$ state via an off-resonant coupling to the $\Done$ level, we induce polarization-dependent light shifts of the Zeeman sublevels, resulting in a tunable polarization-dependent refractive index. For a circularly polarized dressing beam, we observe a rotation of the probe linear polarization, characteristic of the Faraday effect, in the absence of any magnetic field. In addition, for a linearly polarized dressing beam, the probe acquires ellipticity without rotation, corresponding to linear birefringence. More generally, the polarization of the dressing beam controls the axis of rotation of the probe polarization on the Poincaré sphere. Our results establish cold atoms as a versatile platform for engineering and controlling light-induced birefringence and open new perspectives for the fast and reconfigurable control of optical response of resonant media.

\end{abstract}
\maketitle

Birefringence of optical media is widely used to control the polarization of light. Some crystals naturally feature linear birefringence \cite{ferre_linear_1984} and are routinely used in phase plates to control the polarization of light. Birefringence of crystals can be controlled by a static magnetic field \cite{kumari_comprehensive_2022} or by a static electric field \cite{kurtz_physical_1967}, which opened the way to precise control of the polarization of light in technological devices, such as in wave guides \cite{an_active_2019, sahib_elliptical_2022, zafar_recent_2024}, optical fibers \cite{argyros_circular_2009, xiong_complete_2018}, quantum dots \cite{mehdi_giant_2024}, or metasurfaces \cite{hu_all-dielectric_2020, li_metasurface_2024}. In all these devices the polarization state of light is transformed in a unitary way, which can be represented as a rotation of the Poincaré sphere. Two particular types of birefringence can be identified. 1. Linear birefringence: the eigenpolarizations are two orthogonal linear polarization states, and the rotation axis on the Poincaré sphere lies in the equatorial plane. 2. Circular birefringence: the eigenpolarizations are the right- and left-circular polarization states, and the rotation axis coincides with the polar axis. More generally, an arbitrary rotation axis corresponds to elliptical birefringence, whose eigenpolarizations are a pair of orthogonal elliptical polarization states. Having full control of the axis and the angle of rotation in a given system remains a challenge \cite{shi_continuous_2020, sahib_elliptical_2022}.

Another platform where birefringence can occur is dilute media such as atomic gases. There, similar birefringence effects induced by a static magnetic field are harnessed for magnetometry purposes \cite{patton_all-optical_2014,fabricant_how_2023}. In hot vapor experiments, optical control of magneto-optical effects have been reported \cite{pandey_coherent_2008, siddons_optical_2010}, and more interestingly, investigation of light-induced birefringence have been reported \cite{yoon_laser-induced_2004, hu_noiseless_2021}. Cold atoms provide a convenient platform as the inhomogeneous Doppler broadening of the resonance lines is suppressed. The effect of magnetic fields on near-resonant light polarization has been extensively studied \cite{labeyrie_large_2001, pandey_linear_2016, garain_measuring_2024}. This Faraday effect can be used, for example, for nondestructive imaging of atomic clouds \cite{gajdacz_non-destructive_2013, colangelo_simultaneous_2017}. Some theoretical works proposed to mimic the magnetically-induced Faraday effect with an optical field \cite{cho_optically_2005}, but it has not been implemented in cold atomic systems, only signatures of similar effects have been observed in hot atomic vapors (see \cite{yoon_laser-induced_2004, hu_noiseless_2021}).

In this Letter, we demonstrate all-optical control of both the rotation axis and rotation angle of the polarization state of a probe beam propagating through a cold atomic cloud by tuning the polarization, intensity, and frequency of a dressing beam.

More specifically, we create a cold cloud of ytterbium atoms whose birefringence can be engineered using light-induced energy shifts. First, we experimentally demonstrate the independent control of the energies of the $m=\pm 1$ states of the $^3\mathrm{P}_1$ level by optical dressing to the $\Done$ state. When $m = \pm1$ substates of $\Pone$ have different energies, we then show that a linearly-polarized probe tuned at a frequency between these two energies experience optically-induced Faraday rotation. The rotation angle depends on the probe intensity, the energy splitting between the levels, and the optical thickness of the atomic cloud. Finally, we show that the polarization of the dressing beam determines which polarization components of the probe acquire a relative phase shift. This enables arbitrary rotations on the Poincaré sphere for the probe beam. We demonstrate this experimentally by inducing linear birefringence in the atomic cloud and transforming an initially linear polarization of the probe into an elliptical polarization.

\begin{figure}
\includegraphics[width=\linewidth]{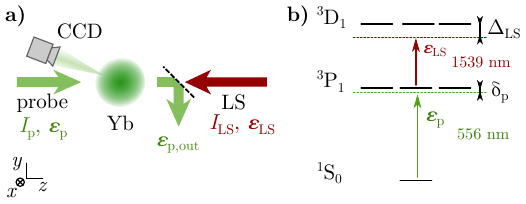} 
\caption{(a) Schematic of the experiment. A probe beam of intensity $I_\mathrm{p}$ and polarization $\bs{\varepsilon}_\mathrm{p}$ illuminates a cold cloud of ytterbium, which is dressed by a counterpropagating light shifter (LS) beam that has intensity $I_\mathrm{LS}$ and polarization $\bs{\varepsilon}_\mathrm{LS}$. The fluorescence is collected on a CCD camera and the transmitted probe beam has polarization $\bs{\varepsilon}_\mathrm{p,out}$ that can be analyzed. (b) Electronic levels and relevant transitions of the ytterbium atoms. The LS beam is detuned by $\Delta_{\mathrm{LS}}$ from the transition $\Pone \rightarrow \Done$, and the probe beam is detuned by $\delta_{\rm p}$ from the transition $\Szero \rightarrow \Pone$.}
\label{Fig1}
\end{figure} 

\subsection{Light shift characterization}

\paragraph{}A schematic of the experimental setup is shown in Fig.~\ref{Fig1}(a): A cold cloud of $^{174}\mathrm{Yb}$ atoms is prepared using a two-stage magneto-optical trapping scheme based on the broad $\Szero \rightarrow {}^{1}\mathrm{P}_1$ transition followed by the narrow $\Szero \rightarrow \Pone$ intercombination line, yielding up to $10^8$ atoms at a temperature of $\sim 15\,\si{\micro\kelvin}$. More details on this preparation can be found in Refs.~\cite{letellier_loading_2023, letellier_piegeage_2024, melo_laser_2024, melo_laser_2024-1}. The cloud has a typical root-mean-square size of $300\,\si{\micro\meter}$ and it is illuminated by a light-shifter beam (LS) with wavelength $\lambda_\mathrm{LS} = 1539\,\si{\nano\meter}$, close to the resonance from the $\Pone$ to $\Done$ states of ytterbium [see Fig.~\ref{Fig1}(b)]. The Gaussian waist of this beam is $w_\mathrm{LS} = 1\,\si{\milli\meter}$, larger than the size of the cloud, and its polarization $\bs{\varepsilon}_\mathrm{LS}$ and intensity $I_\mathrm{LS}$ are control parameters that can be easily changed. This beam is detuned by $|\Delta_{\mathrm{LS}}| \gg \Gamma_{\Done}$, where $\Gamma_{\Done} \approx 2\pi\times 87\,\si{\kilo\hertz}$ is the linewidth of each allowed transition between the Zeeman sublevels of $\Pone$ and $^{3}D_1$ \footnote{We define $\Gamma_{\Done} = \eta |c_{m,m'}|^2/\tau_{\Done}$, where $\tau_{\Done}$ is the total lifetime of the $\Done$ state that has been measured in \cite{beloy_determination_2012}, $\eta = 0.36$ is the branching ratio from the state $\Done$ to the state $\Pone$, and $c_{m,m'}$ is the Clebsch-Gordan coefficient between the Zeeman state $m$ of $\Pone$ and the Zeeman state $m'$ of $\Done$. For such a $J=1$ to $J'=1$ transition, all non-zero Clebsch-Gordan coefficients have an absolute value of $1/\sqrt{2}$.}. This beam induces polarization-dependent light shifts of the Zeeman sublevels.

\begin{figure}
\includegraphics[width=\linewidth]{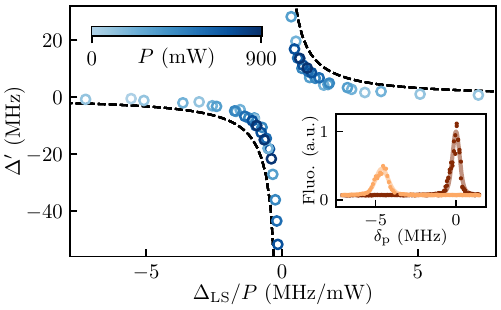} 
\caption{Measurement of the light shift. The light shift $\Delta'$ is measured as the detuning $\Delta_\mathrm{LS}$ and the total power $P$ of the LS beam are varied. The LS beam detuning ranges from $-800$~\si{\mega\hertz} to $800$~\si{\mega\hertz}, and the total power from $80$~\si{\milli\watt} to $700$~\si{\milli\watt}. The result is plotted as a function of $\Delta_\mathrm{LS}/P$, and the power used for each data point is indicated by its color (see color bar). The error bars, smaller than the points size, correspond to the precision with which the center of the spectroscopy line can be extracted, obtained by a bootstrap analysis of the fluorescence datapoints. The dashed line shows the theoretical prediction. An example of the recorded fluorescence signals is shown in the inset: as the probe detuning $\delta_{\rm p}$ is varied, the fluorescence of the center of the cloud is measured when the probe polarization is linear along $x$ (resp. $y$) for the dark-colored curve (resp. light-colored curve). The solid lines are Voigt fits to these measurements.}
\label{Fig2}
\end{figure}

For a circularly polarized LS beam ($\sigma^+$) and a quantization axis chosen along the propagation axis $z$ of the LS beam, the energy of the states $m_z=-1$ and $m_z=0$ is shifted by an amount $\hbar\Delta'$, where
\begin{equation}
\Delta'=-\frac{\Gamma_{\Done}^2}{8\Delta_{\rm LS}}
\frac{I_{\rm LS}}{I_{\rm sat,LS}}.
\label{eq:Delta_LS}
\end{equation}
where $I_\mathrm{LS}$ is the LS intensity and $I_\mathrm{sat,LS} = \pi hc\Gamma_\mathrm{\Done}/(3\lambda_\mathrm{LS}^3)$ the saturation intensity of each allowed transition between the Zeeman substates.

The atomic cloud is then probed with a beam at frequency $\omega_\mathrm{g}$, detuned by $\delta_{\rm p} = \omega_0 - \omega_\mathrm{g}$ with respect to the resonance between $\Szero$ and $\Pone$ (wavelength $\lambda_\mathrm{g} = 556\,\si{\nano\meter}$, natural linewidth $\Gamma_\mathrm{g} = 2\pi\times 182\,\si{\kilo\hertz}$). Its polarization $\bs{\varepsilon}_\mathrm{p}$ and its intensity $I_\mathrm{p}$ will be varied in the following, and its waist $w_\mathrm{p}$ will be adapted for the different experiments presented here.

\begin{figure*}
\includegraphics[width=\linewidth]{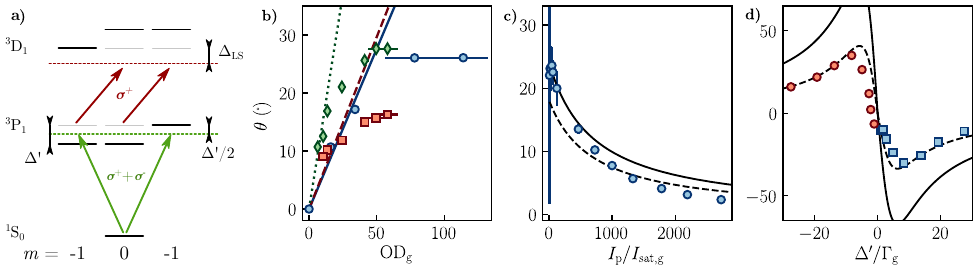} 
\caption{Faraday rotation measurements. (a) Polarization and detunings of the beams : The LS beam has a $\sigma^+$ polarization to shift the $m=0$ and $-1$ states of $\Pone$, and its detuning is $\Delta_{\mathrm{LS}}$. The probe has a linear polarization, which i.e. a $\sigma^+ + \sigma^-$ superposition, and its detuning is $\delta_{\rm p}=\Delta'/2$. (b) Measurement of the rotation angle of an incident linearly-polarized probe as a function of $\mathrm{OD_{g}}$. The three data sets are taken for $(\Delta'/\Gamma_\mathrm{g},\ I_\mathrm{p}/I_\mathrm{sat,g}) = (-33, 30)$ (blue circles), $(-30, 38)$ (red squares) and $(-5.6,38)$ (green diamonds). The respective theoretical predictions are indicated as solid, dashed and dotted lines. (c) Measurement of the rotation angle as a function of $I_\mathrm{p}/I_\mathrm{sat,g}$. The data are taken with a peak $\mathrm{OD_{g}} = 40$ and $\Delta'/\Gamma_g= -26$. The solid line represents the theoretical prediction, which is rescaled by 0.75 to get the dashed line. (d) Measurement of the rotation angle as a function of $\Delta'/\Gamma_g$. The blue squares and red circles are taken respectively with $(\mathrm{OD_{g}},\ I_\mathrm{p}/I_\mathrm{sat,g}) = (45, 32)$ and $(40,32)$. The solid line represents the theoretical prediction, which is rescaled by 0.5 to get the dashed lines. In (b)-(d), the vertical error bars correspond to the noise on the photodiodes used to measure the Stokes parameters, that is propagated to $\theta$.}
\label{Fig3}
\end{figure*}

\paragraph{}Light shifts are the main tool we use to induce birefringence in the atomic cloud. In order to quantify them, we start by measuring the fluorescence of the atomic cloud on a CCD camera (see Fig.~\ref{Fig1}(a)) and vary the probe detuning $\delta_{\rm p}$ to measure the light shift $\Delta'$. The loading of the cloud is shortened to have a relatively low optical depth on the narrow transition, $\mathrm{OD}_\mathrm{g} \approx 2-5$ \cite{glicenstein_situ_2024}. The polarization of the LS beam is set linearly along the $y$ axis, which displaces the state $\ket{+} = (\ket{m_z = -1} + \ket{m_z = +1})/\sqrt{2}$ by $\Delta'$ and keeps the orthogonal state $\ket{-} = (\ket{m_z = -1} - \ket{m_z = +1})/\sqrt{2}$ at its unperturbed energy. The polarization of the probe is then set linearly along $x$ (resp. $y$) to probe the state $\ket{+}$ (resp. $\ket{-}$). The waist of the probe beam is chosen larger than the size of the cloud, but only the fluorescence from the center of the cloud is analyzed \cite{noauthor_see_nodate} to measure the maximum light shift induced by the LS beam and to minimize the impact of inhomogeneous broadening on these measurements. An example of the recorded fluorescence signal is shown on the inset of Fig.~\ref{Fig2}: the signal when $\bs{\varepsilon}_\mathrm{p}$ is linear along $x$ (resp. $y$) is plotted in dark (resp. light) colors and show the unshifted (resp. shifted) transition. The solid lines represent the fits with a Voigt function of these signals. The energy difference $\Delta'$ between the centers of these resonance lines is thus measured and reported in the main graph. The power $P$ of the LS beam and its detuning $\Delta_\mathrm{LS}$ are varied, and $\Delta'$ is plotted as a function of the ratio $\Delta_\mathrm{LS}/P$. The dashed line represents the theoretical prediction from Eq.~(\ref{eq:Delta_LS}), where $I_\mathrm{LS} = 2P/(\pi w_\mathrm{LS}^2)$ is the maximum intensity of the LS beam. The measured light shift follows the theoretical scaling even though it is slightly smaller than the expected value. We attribute the remaining discrepancy to a systematic error in the estimate of the LS beam size.

When setting the LS beam polarization to linear along $x$, the role of the two probed substates is switched and we have verified that the state $\ket{-}$ is displaced by the same light shift. We also have superimposed two LS beams with orthogonal polarizations along $x$ and $y$ and different detunings $\Delta_{\mathrm{LS},x}$ and $\Delta_{\mathrm{LS},y}$ such that they control independently the energy of states $\ket{-}$ and $\ket{+}$ respectively \cite{noauthor_see_nodate}. Note that in the experiment of Fig.~\ref{Fig2}, the third Zeeman state $\ket{m_z = 0}$ is also shifted by $\Delta'$, however, with the geometry used in this experiment, the probe does not couple to this state. More generally, one could independently control the energy of each of the Zeeman states $\ket{-}$, $\ket{0}$ and $\ket{+}$ with three LS beams with respective polarization along $x$, $y$ and $z$, intensities $I_{x,y,z}$ and different detunings $\Delta_{\mathrm{LS},x,y,z}$. The energy shift of the three states would then be
\begin{align}
    \Delta'_{-}&=
    -\frac{\Gamma_{\Done}^2}{8I_{\rm sat,LS}}
    \left(\frac{I_x}{\Delta_{\mathrm{LS},x}}+\frac{I_z}{\Delta_{\mathrm{LS},z}}\right),\\
    \Delta'_{0} &= -\frac{\Gamma^2_{\Done}}{8I_{\mathrm{sat,LS}}}\left(\frac{I_x}{\Delta_{\mathrm{LS},x}} + \frac{I_y}{\Delta_{\mathrm{LS},y}}\right),\\
    \Delta'_{+} &= -\frac{\Gamma^2_{\Done}}{8I_{\mathrm{sat,LS}}}\left(\frac{I_y}{\Delta_{\mathrm{LS},y}} + \frac{I_z}{\Delta_{\mathrm{LS},z}}\right).
\end{align}
As the detunings $\Delta_{\mathrm{LS},x,y,z}$ can independently be chosen to be positive or negative, the three light shifts $\Delta'_{-}$, $\Delta'_0$, and $\Delta'_{+}$ can be independently controlled through the beam intensities and detunings.

As visible in the inset of Fig.~\ref{Fig2}, the shifted resonance is broader than the unshifted one, and its linewidth increases with the light shift $\Delta'$, which we attribute to inhomogeneous broadening due to the finite size of the LS beam and the small angle at which the camera collects the signal: the detected fluorescence originates from atoms experiencing different local light shifts, which broadens the measured lines. We also note that the line broadening $\Gamma_{\mathrm{LS}}$ induced by the dressing laser, scaling as
\begin{equation}
\frac{\Gamma_{\mathrm{LS}}}{\Delta'}
=
\frac{\Gamma_{\Done}}{\Delta_{\mathrm{LS}}},
\end{equation}
remains much smaller than the residual inhomogeneous broadening caused by the nonuniform intensity profile of the LS beam, even within the reduced detection region. Other effects, such as the laser linewidth, Doppler broadening of the atomic cloud, optical-depth effects, and velocity diffusion during fluorescence imaging, are found to be negligible.

\subsection{Faraday-like rotation of the probe polarization}

\paragraph{}We harness these optically controlled light shifts to perform Faraday rotation of a linear probe polarization. The LS beam is circularly polarized, such that the $\ket{m_z=-1}$ state experiences a light shift $\Delta'$, while the $\ket{m_z=+1}$ state remains unshifted. The polarization of the probe is set to linear along $y$: $\bs{\varepsilon}_\mathrm{p} = \bs{u}_y$, and its waist is reduced to $w_{\rm p} = 200\,\si{\micro\meter}$ in order to probe only the center of the cloud, where the integrated optical density is maximal. Its frequency is chosen such that $\delta_{\rm p}= \Delta'/2$ i.e.\ in the middle of the two transitions $\ket{m_z=\pm 1}$ and with $\Delta' \gg \Gamma_\mathrm{g}$, so that the $\sigma^\pm$ components of the probe experience opposite detunings while the residual absorption remains negligible.

The $\sigma^\pm$ components of the probe couple each to the $\ket{m_z=\pm 1}$ state and experience the index of refraction given by $n_\pm = 1+\rho \mathrm{Re}[\alpha(\mp \Delta'/2)]/2$,
where $\alpha(\delta_{\rm p})$ is the polarizability of a two-level atom illuminated with light detuned by $\delta_{\rm p}$ from resonance and $\rho$ is the spatial density of the atoms. This gives:
\begin{equation}
    n_\pm = 1 \mp\frac{3 \rho\lambda_\mathrm{g}^3}{4\pi^2}\frac{\Delta'/(2\Gamma_\mathrm{g})}{1+\frac{\Delta'^2}{\Gamma_\mathrm{g}^2} + \frac{I_\mathrm{p}}{I_\mathrm{sat,g}}}.
\end{equation}

The two components $\sigma^\pm$ of the probe therefore acquire a relative phase shift and the polarization of the probe rotates, analogous to a magnetically-induced Faraday effect.

The rotation angle $\theta$ of the probe polarization after propagation through the cloud is then given by: $\theta = \pi\int \D z\,[n_-(z) - n_+(z)]/\lambda_\mathrm{g}$,
and if the probe intensity is constant through the cloud, we obtain:
\begin{equation}\label{eq:theta}
    \theta = \frac{\mathrm{OD}_\mathrm{g}}{2}\frac{\Delta'/\Gamma_\mathrm{g}}{1+\frac{\Delta'^2}{\Gamma_\mathrm{g}^2} + \frac{I_\mathrm{p}}{I_\mathrm{sat,g}}},
\end{equation}
where $\mathrm{OD}_\mathrm{g} = 3\lambda_\mathrm{g}^2\int \D z\,\rho(z)/(2\pi)$ is the on-resonance optical depth of the cloud. This formula is valid when the probe is not absorbed by the cloud, which requires the detuning $\Delta'/2$ to be larger than $\Gamma_\mathrm{g}$. The experiments reported below are performed in this regime.

Experimentally, we probe this effect by measuring the rotation angle $\theta$ and varying the three experimentally tunable parameters $\mathrm{OD}_\mathrm{g}$, $I_\mathrm{p}$ and $\Delta'$. We send a linearly polarized probe through the dressed atomic cloud. Due to the non-uniform density distribution of the atomic cloud, our theoretical predictions include a correction for the finite size of the probe \cite{noauthor_see_nodate}. The transmitted polarization is analyzed using a polarimetric detection scheme where the Stokes parameters $s_{1,2,3}$ that characterize the polarization are measured \cite{noauthor_see_nodate}, from which the rotation angle $\theta$ is extracted.

In the first experiment [Fig.~\ref{Fig3}(b)], we choose three sets of parameters for the LS beam ($\Delta_\mathrm{LS}$, $P$) and increase the optical depth of the atomic cloud by varying the number of atoms loaded in the trap. We observe the expected Faraday rotation, and we see that at low optical depth, the measurements follow the theoretical prediction of Eq.~(\ref{eq:theta}) represented by the lines in Fig.~\ref{Fig3}(b). For larger OD, the rotation angle then departs from this prediction. We attribute this deviation to the difficulty in the measurement of such large optical densities. We verified that we don't observe any depolarization effect nor any significant absorption of the light for these measurements.

In a second experiment [Fig.~\ref{Fig3}(c)], we fix the optical depth of the cloud ($\mathrm{OD}_\mathrm{g} = 40$) and the parameters of the LS beam, and increase the intensity of the probe beam. As the saturation parameter $I_{\rm p}/I_{\rm sat,g}$ increases, the rotation effect is reduced, as expected from Eq.~(\ref{eq:theta}). The theoretical prediction is plotted as a solid line. Since we have a large optical density, we rescale it by a factor of 0.75 (dashed line) to take into account the reduced rotation observed in the previous experiment. With our setup, the signal-to-noise ratio of the photodiode measurement limits our ability to probe the low saturation regime and the smallest intensity we report here is $I_\mathrm{p}/I_\mathrm{sat,g} = 8$.

In a third experiment [Fig.~\ref{Fig3}(d)], we vary the energy shift $\Delta'$. In a first set of data we keep the laser detuning $\Delta_\mathrm{LS}$ fixed and vary the intensity $I_\mathrm{LS}$ of the LS (red circles), and in a second set of data we change the sign of $\Delta_\mathrm{LS}$ and vary $I_\mathrm{LS}$ (blue squares), where we also observe that the direction of rotation is inverted. The theoretical prediction is plotted as a solid line and we rescale it by a factor 0.5 to compare it with our measurements.

These experiments demonstrate that our atomic sample, dressed with a $\sigma^+$ light shifter beam, acts on the probe linear polarization, when its frequency is well-chosen, as a Faraday rotator with a controlled rotation angle. More generally, the action of the atomic cloud on an arbitrary probe polarization is represented on the Poincaré sphere as a rotation by an angle $\theta$ about the axis defined by the $\sigma^+$ and $\sigma^-$ polarization states.

\subsection{Elliptical birefringence and arbitrary Poincaré sphere rotations}

\begin{figure}
\includegraphics[width=\linewidth]{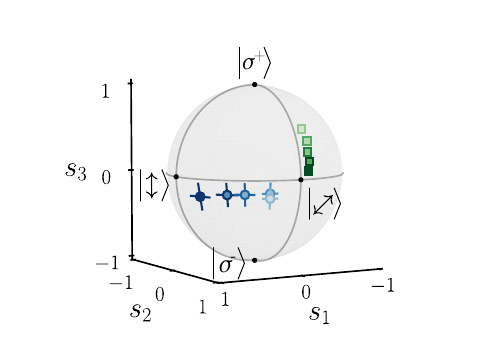} 
\caption{Generalized Faraday rotations on the Poincaré sphere. Blue circles: same data as in Fig.~\ref{Fig3}(b), now represented on the Poincaré sphere. The LS beam is $\sigma^+$ polarized. The probe is initially linearly polarized (measured in the absence of atoms with the darkest blue circle), and its polarization as $\mathrm{OD}_{\rm g}$ increases is represented with the lighter shades of blue. Green squares: the LS beam is linearly polarized along $y$. The probe is initially linearly polarized at $45^\circ$ (darkest square) and as the optical depth increases, we observe that the polarization becomes elliptical (squares in shades of green). The experimental parameters for these last datapoints are: $(\Delta'/\Gamma_\mathrm{g}, I_\mathrm{p}/I_\mathrm{sat,g}) = (-24, 180)$, and the optical densities of the cloud range from 10 to 30. The error bars correspond to the noise on the photodiodes used to measure the Stokes parameters.}
\label{Fig4}
\end{figure}

This optically-induced Faraday rotation of the probe polarization can be further generalized by changing the polarization of the LS beam to induce elliptical birefringence in the atomic cloud (see \cite{noauthor_see_nodate} for more details) and control the rotation axis on the Poincaré sphere. When it is set to $\bs{\varepsilon}_\mathrm{LS} = \cos(\vartheta/2)\bs{\sigma}^+ + \sin(\vartheta/2)\E^{\I\varphi}\bs{\sigma}^-$, the shifted state $\ket{\psi_-}$ and the unshifted state $\ket{\psi_+}$ are now linear combinations of the states $\ket{m_z = \pm 1}$ of the $\Pone$ manifold:
\begin{align}
    \ket{\psi_+} &= \cos(\vartheta/2)\ket{m_z = 1} + \sin(\vartheta/2)\E^{\I\varphi}\ket{m_z = -1},\\
    \ket{\psi_-} &= - \sin(\vartheta/2)\E^{\I\varphi}\ket{m_z = 1} + \cos(\vartheta/2)\ket{m_z = -1}.
\end{align}
A probe beam with its frequency midway between the resonance frequencies of these two dressed states then decomposes along the components parallel and orthogonal to the polarization of the LS beam, and these components acquire a relative phase shift while propagating through the atomic cloud. This corresponds to a rotation on the Poincaré sphere about the axis defined by the LS polarization and its orthogonal counterpart, through an angle $\theta$ given by Eq.~(\ref{eq:theta}).
We demonstrate this ability to choose the direction of the axis of rotation of the probe polarization by setting $\bs{\varepsilon}_{\rm LS}$ to be linearly polarized along the vertical direction, thereby inducing linear birefringence. We perform the same experiment presented above, this time with a probe having a linear polarization at $45\si{\degree}$ from the vertical. The probe is thus an equal superposition of vertical and horizontal components that are initially in phase. As the probe passes through the atomic cloud, these two components dephase and we expect the probe polarization to become elliptical. The probe polarization after propagation through the cloud is shown in Fig.~\ref{Fig4} (green squares), where the three Stokes parameters are plotted on the Poincaré sphere. The initial polarization (filled dark square) is linear and at $45\si{\degree}$ and as the optical density of the cloud is increased, the polarization rotates along a meridian of the sphere to become more elliptical. As a comparison, one of the data presented on Fig.~\ref{Fig3}(b) is shown as blue circles: the initial polarization is close to a vertical linear polarization, and as the optical density of the cloud is increased, the polarization stays linear and rotates on the equator of the sphere.

\paragraph{}In conclusion, we experimentally demonstrated that an optical dressing of atomic excited states can replace a static magnetic field to realize an optically-induced Faraday effect with a cloud of cold atoms. Such a dressing breaks time-reversal symmetry and could therefore be used to realise an equivalent of an optical isolator. We also demonstrated that controlling the polarization of the optical dressing harnesses the richness of the tensor light-shift to choose the axis of rotation on the Poincaré sphere at will. In particular, a probe beam with linear polarization can become elliptically polarized after passing through the cloud.
Such an effect can be obtained by applying a static electric field, but the simplicity of our control with the dressing polarization makes it more elegant and experimentally simpler. The polarization degree of freedom can be exploited to engineer spatially-structured polarization rotations: by dressing the atomic cloud with a non-uniform beam or with a vector light field one can engineer complex polarization states of light \cite{wang_vectorial_2020}. Furthermore, the dependence of the rotation angle on the probe intensity, combined with our ability to modulate this intensity on sub-microsecond timescales, enables the generation of rapidly varying polarization states of the probe beam \cite{shen_roadmap_2023,jaffray_all-optical_2026}. Such control could find applications in the study of time-dependent mesoscopic phenomena in cold atomic systems \cite{labeyrie_hanle_2002, sigwarth_magnetic_2004}.

\paragraph{Acknowledgements:} This work was performed in the framework of the European project ANDLICA (ERC Advanced grant No. 832219), the French National Research Agency (projects PACE-IN (ANR19-QUAN-003), LiLoA (ANR23-CE30-0035) and QUTISYM (ANR-23-PETQ-0002)). D.B.O is supported by European Union’s Horizon 2020 research and innovation program under the Marie Sk\l odowska-Curie grant agreement No. 101105291.

\bibliography{biblio.bib}

\cleardoublepage
\setcounter{equation}{0}
\setcounter{figure}{0}
\renewcommand{\theequation}{S\arabic{equation}}
\renewcommand{\thefigure}{S\arabic{figure}}

\appendix
\onecolumngrid 

\begin{center}
{\large \textbf{Supplementary Material to:}}\\[0.5em]
{\large \textbf{All-Optical Control of Birefringence in a Cold Atomic Ensemble
}}
\end{center}

\twocolumngrid

\section{Measuring the light shift}

\begin{figure}
\includegraphics[width=\linewidth]{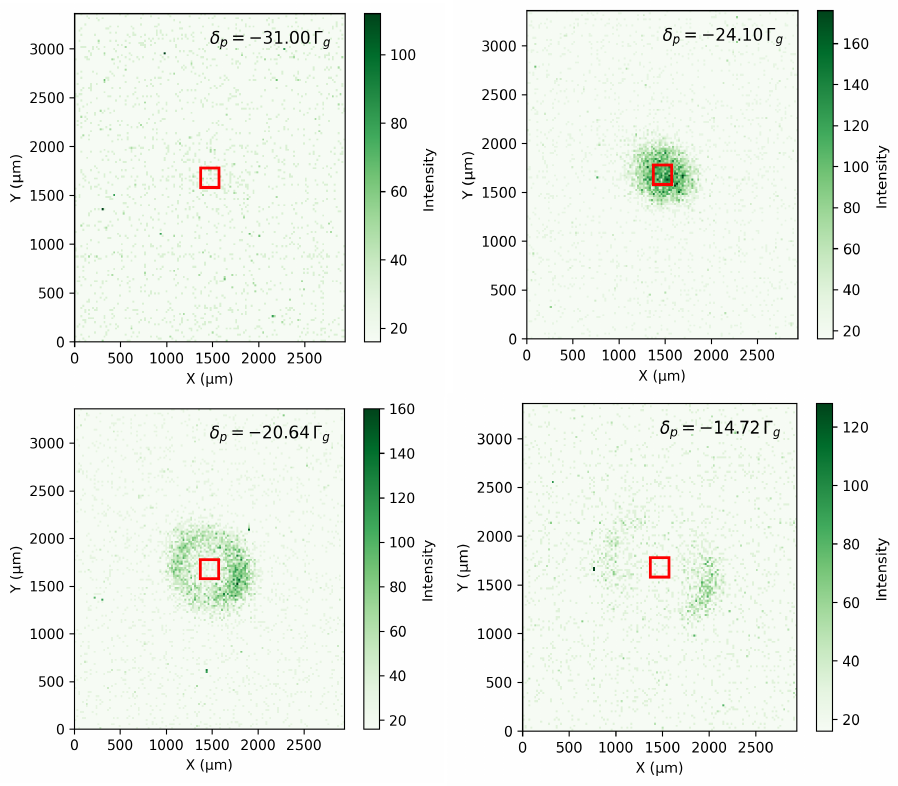} 
\caption{Measuring the light shift. The four panels present examples of fluorescence pictures taken at different probe detunings $\delta_{\rm p}$ relative to the unperturbed $\Szero\rightarrow \Pone$ transition, indicated at the top of each frame. When the probe beam is far detuned (frame a), the atomic fluorescence is negligible and the frame is empty. When the probe beam is resonant with the maximally-shifted atomic transition in the center of the cloud, only this central region reemits photons (frame b). When detuning the probe beam less and less (frames c and d), different regions of the cloud get on resonance as they experience smaller light shifts due to the inhomogeneous distribution of the light-shifter intensity. To measure the maximum light shift, only the central region (in red) is analyzed.}
\label{FigS4}
\end{figure}

The light shift of the $\Pone$ excited state is measured by fluorescence spectroscopy of the atomic cloud.
To minimize multiple scattering effects, the resonant optical depth on the probe transition is maintained in the range $\mathrm{OD}_{g} \approx 2$--$5$.

The light-shifter (LS) beam at $1539\,\si{\nano\meter}$, detuned by $\Delta_{\rm LS}$ with respect to the $\Pone \rightarrow \Done$ transition, 
propagates along the $-z$ direction on the cloud.
The beam has a Gaussian waist $w_{\mathrm{LS}} \simeq 1\,\si{\milli\meter}$, larger than the atomic cloud, ensuring a nearly uniform intensity over the region of interest. Its intensity $I_\mathrm{LS}$ and polarization $\varepsilon_{\mathrm{LS}}$ are independently controlled.
The detuning is chosen such that $|\Delta_{\rm LS}| \gg \Gamma_{\Done}$, so that the population of the $\Done$ state can be neglected.

The induced light shift is measured by scanning the frequency of a weak probe beam across the $\Szero \rightarrow \Pone$ transition. The probe beam uniformly illuminates the cloud along the $+z$ direction, 
and its polarization is chosen to address selected superpositions of Zeeman sublevels of the $\Pone$ state.

The LS and probe beams are applied simultaneously for a duration of $1\,\si{\milli\second}$, during which atomic fluorescence is recorded on a CCD camera positioned at an angle of approximately $10^{\circ}$ from the backward direction of the probe beam.

For each probe detuning, the fluorescence of the atomic cloud is imaged on a CCD camera (see Figure \ref{FigS4}). The spectrum is constructed by summing the fluorescence counts within a $210\,\si{\micro\meter} \times 210\,\si{\micro\meter}$ square region centered on the LS beam axis indicated in red in Figure \ref{FigS4}, corresponding to the region of maximum and most homogeneous light shift. Over this region, the Gaussian intensity profile of the LS beam varies by less than $3\%$.

For each set of experimental parameters, fluorescence spectra are recorded over multiple independent experimental realizations. The resonance position is extracted by fitting each spectrum with a Voigt profile, consisting of a Lorentzian of fixed width $\Gamma_{g}$ convolved with a Gaussian accounting for Doppler and inhomogeneous broadening.  The center frequency of the fitted resonance yields the light-shifted transition frequency. By comparing spectra acquired for different probe polarizations, the differential light shift between dressed and undressed states is extracted.

Statistical uncertainties on the light shift are estimated using a bootstrap resampling method, in which the datapoints of a spectrum are resampled with replacement and refitted to obtain several estimates of the resonance frequency. The statistical uncertainty on the reported resonance frequency is taken as the standard deviation of the bootstrap estimates.

\section{Independent control of the energy of two Zeeman substates}

\begin{figure}
\includegraphics[width=\linewidth]{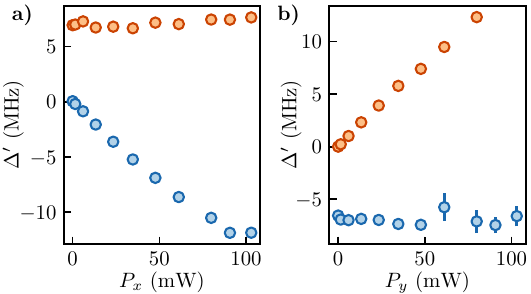} 
\caption{Independent control of the energy of two Zeeman substates. Two orthogonally polarized light shifter beams are sent on the atomic cloud with different detunings and independently-tunable powers. The light shift of the two substates $\ket{+}$ and $\ket{-}$ are measured and plotted as orange and blue circles respectively. In panel a, the power $P_x$ of one light shifter beam is varied while the other $P_y$ is fixed, and in panel b, $P_x$ is fixed and $P_y$ is varied. The error bars correspond to the precision with which the center of the spectroscopy line can be extracted, obtined by a bootstrap analysis of the datapoints.}
\label{FigS3}
\end{figure} 

We demonstrate independent optical control of the energies of the substates $\ket{+}$ and $\ket{-}$ using two light-shift beams with orthogonal linear polarizations along $x$ and $y$, and detuned from the $\Pone \rightarrow \Done$ transition by  $\Delta_{\rm LS} = -76$ and $+84\,\si{\mega\hertz}$, respectively.

In a first experiment, the total power $P_y$ of the $y$-polarized beam is fixed, while the power $P_x$ of the $x$-polarized beam is varied. For each value of $P_y$, two resonance curves are measured, one with an $x$-polarized and one with a $y$-polarized probe beam. For each of them, the center of the resonance is extracted and reported on Figure \ref{FigS3}a. The light shift induced by the $x$-polarized (resp. $y$-polarized) light shifter beam is probed with the $y$-polarized (resp. $x$-polarized) probe and the corresponding shift with respect to the unshifted line is plotted in blue (resp. orange). We observe that the light shift of the $\ket{+}$ state, induced by the $y$ component of the light shifter beams is kept constant when increasing the power of the orthogonal polarization, while the light shift of the $\ket{-}$ state is linearly increasing with this power. Note that here the detunings of the two light shifter beams have been chosen here with opposite sign, which leads to a positive (resp. negative) light shift for the state $\ket{+}$ (resp. $\ket{-}$). In a second experiment, $P_x$ is kept constant while $P_y$, demonstrating the complementary tunability in Figure \ref{FigS3}b.

These experiments demonstrate the independent optical control of the energy of the two substates $\ket{+}$ and $\ket{-}$.

\section{Measurement of Stokes Parameters}

\begin{figure}
\includegraphics[width=0.7\linewidth]{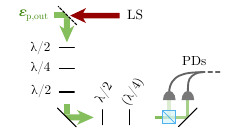} 
\caption{Stokes polarimetry. The transmitted probe with polarization $\varepsilon_{\mathrm{p,out}}$ is separated from the LS beam with a dichroic mirror. This induces a small change of polarization that is corrected with the three following waveplates. The two next waveplates choose the measurement basis of the polarization, and the polarizing beam splitter (PBS) and the two photodiodes (PDs) perform this measurement.}
\label{FigS5}
\end{figure}

The polarization state of the transmitted probe beam is characterized using full Stokes polarimetry following the formalism of Ref.~\cite{labeyrie_large_2001}.
We first compensate the birefringence induced by the dichroic plate that separates the probe beam from the light shifter beam (see figure \ref{FigS5}). For this we use a series of $\lambda/2 - \lambda/4 - \lambda/2$ plates. The probe beam then passes through one or two waveplates that selects the measurement basis, followed by a polarizing beam splitter that projects the initial polarization on the chosen basis. The intensities at the two PBS outputs are recorded on photodiodes.

Measurements are performed sequentially in three polarization bases:
\begin{enumerate}
    \itemsep=0pt
    \item{linearly polarized and vertical/horizontal (one $\lambda/2$ waveplate): we measure $I_\parallel$ and $I_\perp$,}
    \item{linearly polarized and at $\pm45^\circ$ (one $\lambda/2$ waveplate): we measure $I_{45\circ}$ and $I_{-45\circ}$,}
    \item{circularly polarized, with right/left chirality (one $\lambda/2$ and one $\lambda/4$ waveplate): we measure $I_{\sigma+}$ and $I_{\sigma-}$.}
\end{enumerate}
These measurements allow reconstruction of the full Stokes vector of the transmitted probe.

\paragraph{Data Acquisition and Timing Window:}

For each value of the light shift $\Delta'$, the probe frequency is set midway between the shifted and unshifted resonances, $\delta_{\rm p} = \Delta'/2$, maximizing the birefringence while minimizing the absorption.

For each experimental cycle, the probe signal is recorded for $1~\si{\milli\second}$.
Only the interval between $t = 70~\si{\micro\second}$ and $t = 270~\si{\micro\second}$ is retained for analysis.
The initial $70~\si{\micro\second}$ contain transient electronic spikes and optical switching artifacts associated with the AOM driver that turns on the probe beam.
Restricting the analysis to a $200~\si{\micro\second}$ window ensures a steady-state signal with good signal-to-noise ratio.

Each measurement is repeated ten times for a given set of parameters.
An additional background signal is recorded with the probe blocked and subtracted from all detector signals.

\paragraph{Extraction of Stokes Parameters and Rotation Angle:}

After background subtraction, the total probe intensity is defined as
\begin{equation}
s_0 = I_\parallel + I_\perp .
\end{equation}

The normalized Stokes parameters are then computed as
\begin{align}
s_1 &= \frac{I_\parallel - I_\perp}{s_0}, \\
s_2 &= \frac{2 I_{45\circ}}{s_0} - 1, \\
s_3 &= \frac{2 I_{\sigma+}}{s_0} - 1.
\end{align}

Normalization by $s_0$ suppresses common-mode probe power fluctuations.
The Faraday rotation angle $\theta$, defined as the orientation of the major axis of the polarization ellipse relative
to the incident polarization, is extracted from
\begin{equation}
\tan(2\theta) = \frac{s_2}{s_1}.
\end{equation}

The ellipticity of the transmitted probe can additionally be quantified as
\begin{equation}
\epsilon =
\frac{1}{2}
\arcsin\!\left(
\frac{s_3}{\sqrt{s_1^2 + s_2^2 + s_3^2}}
\right).
\end{equation}

\paragraph{Uncertainty Estimation}

Statistical uncertainties on the Stokes parameters are obtained from the standard deviation of repeated measurements.
The uncertainty on the extracted rotation angle $\theta$ is computed using Gaussian error propagation,
\begin{equation}
\sigma_\theta^2 =
\left(
\frac{\partial \theta}{\partial s_1} \sigma_{s_1}
\right)^2
+
\left(
\frac{\partial \theta}{\partial s_2} \sigma_{s_2}
\right)^2,
\end{equation}
with
\begin{align}
\frac{\partial \theta}{\partial s_1} &= -\frac{s_2}{2(s_1^2 + s_2^2)}, \\
\frac{\partial \theta}{\partial s_2} &= \phantom{-}\frac{s_1}{2(s_1^2 + s_2^2)} .
\end{align}

\paragraph{Optical Depth and Mode-Overlap Correction}

The optical depth relevant for the green probe transition is obtained by rescaling the optical depth measured on the strong blue $\Szero\rightarrow \Pone$ transition,
\begin{equation}
\mathrm{OD}_\mathrm{g} = \eta\,\mathrm{OD}_{\mathrm{blue}}, \qquad \eta = 1.7.
\end{equation}
This factor $\eta$ comes both from the ratio of the scattering cross-sections $(\lambda_g/\lambda_b)^2$ and from the Doppler effect due to the non-zero temperature of the atoms (see \cite{glicenstein_situ_2024} for more details).

Because the probe samples only part of the Gaussian atomic cloud, the measured rotation corresponds to an intensity-weighted average over the probe mode.
The effective optical depth used to determine the theoretical rotation is therefore reduced to
\begin{equation}
\mathrm{OD}_{\mathrm{eff}} =
\frac{R^2}{R^2 + w_\mathrm{p}^2}\,
\mathrm{OD}_\mathrm{g},
\end{equation}
where $R$ is the root mean square radius of the cloud.
For $w_p \simeq 200~\mu\mathrm{m}$ and $R \simeq 250~\mu\mathrm{m}$, this yields $\mathrm{OD}_{\mathrm{eff}} \approx 0.6\,\mathrm{OD}_\mathrm{g}$.

\section{Generalization of Faraday rotations}

\begin{figure}
\includegraphics[width=\linewidth]{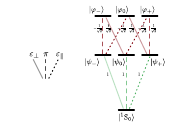} 
\caption{Coupling diagram between the three manifolds $\Szero$, $\Pone$ and $\Done$ adapted to the photon polarization basis $\{\hat{\varepsilon}_\parallel, \hat{\varepsilon}_\perp, \hat{\pi}\}$. In this reference frame, the coupling diagram and the Clebsch-Gordan coefficients are exactly the same as in the initial reference frame associated with the photon polarization basis $\{\hat{\sigma}^+, \hat{\sigma}^-, \hat{\pi}\}$.}
\label{FigS1}
\end{figure}

The optically-induced Faraday effect originates from polarization-dependent light shifts that break the degeneracy of Zeeman sublevels. To describe this mechanism, we express the light–atom coupling in a basis adapted to the polarization of the dressing field. This approach naturally reveals the emergence of elliptical birefringence for arbitrary polarization states.

In the basis $\{\ket{\Pone, m_z},\ket{\Done, m_z}\}$, the light couplings are written as:
\begin{align}
    \hat{\sigma}^+_\mathrm{LS} &= -\frac{1}{\sqrt{2}}\left(\ketbra{m'_z = 0}{m_z = -1} + \ketbra{m'_z = 1}{m_z = 0}\right) + \mathrm{c.c.},\\
    \hat{\pi}_\mathrm{LS} &= \frac{1}{\sqrt{2}}\left(-\ketbra{m'_z = -1}{m_z = -1} +\ketbra{m'_z = 1}{m_z = 1}\right) + \mathrm{c.c.},\\
    \hat{\sigma}^-_\mathrm{LS} &= \frac{1}{\sqrt{2}}\left(\ketbra{m'_z = -1}{m_z = 0} + \ketbra{m'_z = 0}{m_z = 1}\right) + \mathrm{c.c.},
\end{align}
where we have used the notations $\ket{m'_z} \equiv \ket{\Done, m_z}$ and $\ket{m_z} \equiv \ket{\Pone, m_z}$.

We now suppose that the polarization of the LS beam is $\ket{\varepsilon_\parallel} = \cos(\vartheta/2)\ket{\sigma^+} + \sin(\vartheta/2)\E^{\I\varphi}\ket{\sigma^-}$, where the parameters $\vartheta,\varphi$ parametrize the Poincaré sphere of the LS beam polarization. The coupling is thus given by
\begin{align}
    \hat{\varepsilon}_{\mathrm{LS},\parallel} &= \frac{1}{\sqrt{2}}\Big\{\cos(\vartheta/2)\left[-\ketbra{m'_z = 0}{m_z = -1} - \ketbra{m'_z = 1}{m_z = 0}\right]\nonumber\\
    &\quad + \sin(\vartheta/2)\E^{\I\varphi}\left[\ketbra{m'_z = -1}{m_z = 0} + \ketbra{m'_z = 0}{m_z = 1}\right]\Big\}\nonumber\\
    &\quad + \mathrm{c.c.}
\end{align}

We perform a change of basis in the $\Pone$ and $\Done$ manifolds by defining the states
\begin{align}
    \ket{\psi_+} &= \cos(\vartheta/2)\ket{m_z = 1} + \sin(\vartheta/2)\E^{\I\varphi}\ket{m_z = -1},\nonumber\\
    \ket{\psi_0} &= \ket{m_z = 0},\nonumber\\
    \ket{\psi_-} &= - \sin(\vartheta/2)\E^{-\I\varphi}\ket{m_z = 1} + \cos(\vartheta/2)\ket{m_z = -1},
\end{align}
and
\begin{align}
    \ket{\phi_+} &= \cos(\vartheta/2)\ket{m'_z = 1} - \sin(\vartheta/2)\E^{\I\varphi}\ket{m'_z = -1},\\
    \ket{\phi_0} &= \ket{m'_z = 0},\\
    \ket{\phi_-} &= \sin(\vartheta/2)\E^{-\I\varphi}\ket{m'_z = 1} + \cos(\vartheta/2)\ket{m'_z = -1}.
\end{align}
The inverse transformations read:
\begin{align}
    \ket{m_z = 1} &= \cos(\vartheta/2)\ket{\psi_+} - \sin(\vartheta/2)\E^{\I\varphi}\ket{\psi_-},\\
    \ket{m_z = 0} &= \ket{\psi_0},\\
    \ket{m_z = -1} &= \sin(\vartheta/2)\E^{-\I\varphi}\ket{\psi_+} + \cos(\vartheta/2)\ket{\psi_-},
\end{align}
and
\begin{align}
    \ket{m'_z = 1} &= \cos(\vartheta/2)\ket{\phi_+} + \sin(\vartheta/2)\E^{\I\varphi}\ket{\phi_-},\\
    \ket{m'_z = 0} &= \ket{\phi_0},\\
    \ket{m'_z = -1} &= -\sin(\vartheta/2)\E^{-\I\varphi}\ket{\phi_+} + \cos(\vartheta/2)\ket{\phi_-},
\end{align}
such that the coupling of the light shifter polarization becomes in this new basis:
\begin{equation}
\hat{\varepsilon}_{\mathrm{LS},\parallel} = -\frac{1}{\sqrt{2}}\ketbra{\phi_0}{\psi_-} - \frac{1}{\sqrt{2}}\ketbra{\phi_+}{\psi_0} + \mathrm{c.c.}
\end{equation}

We can also define the orthogonal polarization of the light shifter beam:
\begin{equation}
    \ket{\varepsilon_{\mathrm{LS},\perp}} = -\sin(\vartheta/2)\E^{-\I\varphi}\ket{\sigma^+} + \cos(\vartheta/2)\ket{\sigma^-},
\end{equation}
which provides the coupling, expressed in the new basis:
\begin{equation}
    \hat{\varepsilon}_{\mathrm{LS},\perp} = \frac{1}{\sqrt{2}}\ketbra{\phi_0}{\psi_+} + \frac{1}{\sqrt{2}}\ketbra{\phi_-}{\psi_0} + \mathrm{c.c}.
\end{equation}
And we can finally express the $\hat{\pi}$ polarization:
\begin{equation}
    \hat{\pi}_\mathrm{LS} = -\frac{1}{\sqrt{2}}\ketbra{m'=-1}{m=-1} + \frac{1}{\sqrt{2}}\ketbra{m'=1}{m=1} + \mathrm{c.c.},
\end{equation}
which in the new basis reads:
\begin{equation}
    \hat{\pi} = -\frac{1}{\sqrt{2}}\ketbra{\phi_-}{\psi_-} + \frac{1}{\sqrt{2}}\ketbra{\phi_+}{\psi_+} + \mathrm{c.c.}
\end{equation}
We can also perform the same change of basis to describe the coupling from the ground state $\Szero$ to the $\Pone$ manifold: the three polarizations states $\ket{\varepsilon_\parallel}, \ket{\pi}, \ket{\varepsilon_\perp}$ induce the following couplings between the ground state noted $\ket{\Szero}$ and the states $\ket{\psi_{\pm,0}}$ are:
\begin{align}
    \hat{\varepsilon}_{\mathrm{g},\parallel} &= \ketbra{\psi_+}{\Szero} + \mathrm{c.c},\\
    \hat{\pi}_\mathrm{g} &= \ketbra{\psi_0}{\Szero} + \mathrm{c.c},\\
    \hat{\varepsilon}_{\mathrm{g},\perp} &= \ketbra{\psi_-}{\Szero} + \mathrm{c.c}.
\end{align}

In the new basis, we end up with the coupling scheme represented on figure \ref{FigS1}. This justifies the fact that, in the case where $\varepsilon_\mathrm{LS}$ is linear and vertical, the proper basis for the $\Pone$ manifold is the basis $(\ket{m_z = -1} + \ket{m_z = +1})/\sqrt{2}$, $\ket{m_z = 0}$, $(\ket{m_z = -1} - \ket{m_z = +1})/\sqrt{2}$, that can be probed respectively with the probe linearly polarized in the $x$, $z$ and $y$ direction.

\paragraph{}This allows us to understand the effect of a light-shifter beam with arbitrary polarization $\ket{\varepsilon_\parallel}$: it displaces the two states $\ket{\psi_-}$ and $\ket{\psi_0}$ by the same amount $\Delta_\mathrm{LS}$. If now the probe beam is sent on the cloud with a detuning $\Delta_\mathrm{LS}/2$ and a polarization $\ket{\varepsilon_\mathrm{g}}$ that we decompose on the proper basis:
\begin{equation}
    \ket{\varepsilon_\mathrm{g}} = \alpha\ket{\varepsilon_\parallel} + \beta\ket{\varepsilon_\perp},
\end{equation}
then this polarization comes out of the cloud with polarization 
\begin{equation}
    \ket{\varepsilon_{\mathrm{g},\mathrm{out}}} = \alpha\E^{\I\theta_\mathrm{int}}\ket{\varepsilon_\parallel} + \beta\E^{-\I\theta_\mathrm{int}}\ket{\varepsilon_\perp},
\end{equation}
where $\theta_\mathrm{int}$ is the rotation angle defined in equation (7) of the main text. This transformation describes a rotation of angle $\theta_\mathrm{int}$ around the axis passing through the states $\ket{\varepsilon_{\parallel}}$ and $\ket{\varepsilon_{\perp}}$ on the Poincaré sphere.

\begin{figure}
\includegraphics[width=\linewidth]{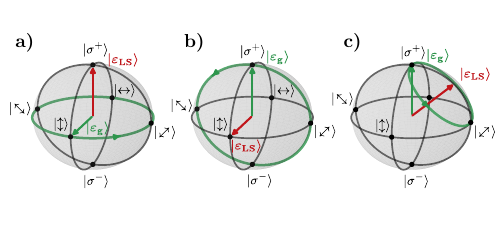} 
\caption{Rotation of the polarization $\varepsilon_\mathrm{g}$ as the polarization of the light shifter is set to $\varepsilon_\mathrm{LS}$, in three cases: a) $\varepsilon_\mathrm{LS} = \sigma^+$, and the polarization $\varepsilon_\mathrm{g}$ rotates as for a Faraday rotation. b) $\varepsilon_\mathrm{LS}$ is linear and vertical, $\varepsilon_\mathrm{g}$ is initially circular and oscillates between linear at $45\si{\degree}$ and circular as $\theta_\mathrm{int}$ increases. c) A more general case where $\varepsilon_\mathrm{LS}$ is pointing in an arbitrary direction and $\varepsilon_\mathrm{g}$ is rotating around it.}
\label{FigS2}
\end{figure} 

This reasoning can be applied in the two particular cases, illustrated in figure \ref{FigS2}:
\begin{itemize}
    \item{$\ket{\varepsilon_\mathrm{LS}} = \ket{\sigma^+}$, which means that $\vartheta = 0$. The proper basis of the $\Pone$ and $\Done$ manifolds is the $\ket{m_z}$ basis, and a linear initial probe with $\ket{\varepsilon_\mathrm{g}} = (\ket{\sigma^+} + \ket{\sigma^-})/\sqrt{2}$ stays linear and the direction of this polarization rotates (see Fig \ref{FigS2}a).
    This corresponds to the standard Faraday rotation mechanism, realized here without any external magnetic field.}
    \item{$\ket{\varepsilon_\mathrm{LS}} = (\ket{\sigma^+} + \ket{\sigma^-})/\sqrt{2}$, which means that $\vartheta = \pi/2$ and $\varphi = 0$. This corresponds to a linear vertical polarization. Then if $\ket{\varepsilon_\mathrm{g}} = (\ket{\sigma^+} + \I\ket{\sigma^-})/\sqrt{2}$, i.e. linear at $45\si{\degree}$, the output polarization $\ket{\varepsilon_\mathrm{g,out}}$ will oscillate between a linear and circular polarization as $\theta_\mathrm{int}$ increases (see Fig \ref{FigS2}b).}
\end{itemize}
The last schematics (Fig \ref{FigS2}c) illustrates a more general case where the polarization of the light shifter is elliptical and the initial probe polarization is arbitrarily chosen as linear and at 45°.

\end{document}